\documentclass[preprint,12pt]{elsarticle}

\usepackage{setspace} 
\usepackage{amsmath, amssymb}
\usepackage{bm}
\usepackage{graphicx}
\usepackage{subfig}
\usepackage{gensymb}
\usepackage{algorithmic}
\usepackage{algorithm}
\usepackage{color}

\bmdefine{\bA}{A}
\bmdefine{\ba}{a}
\bmdefine{\bB}{B}
\bmdefine{\bQ}{Q}
\bmdefine{\bE}{E}
\bmdefine{\bF}{F}
\bmdefine{\bG}{G}
\bmdefine{\bD}{D}
\bmdefine{\bd}{d}
\bmdefine{\bc}{c}
\bmdefine{\bg}{g}
\bmdefine{\bX}{X}
\bmdefine{\bx}{x}
\bmdefine{\bL}{L}
\bmdefine{\bJ}{J}
\bmdefine{\btau}{\tau}
\bmdefine{\bn}{n}
\bmdefine{\bq}{q}
\bmdefine{\bt}{t}
\bmdefine{\bb}{b}
\bmdefine{\bv}{v}
\bmdefine{\bu}{u}
\bmdefine{\bU}{U}
\bmdefine{\bsigma}{\sigma}
\bmdefine{\br}{r}
\bmdefine{\bS}{S}
\bmdefine{\bT}{T}
\bmdefine{\bP}{P}
\bmdefine{\bR}{R}
\bmdefine{\bV}{V}
\bmdefine{\bW}{W}
\bmdefine{\bhE}{\hat{E}}
\bmdefine{\bzeta}{\zeta}
\bmdefine{\bphi}{\phi}
\bmdefine{\bsigma}{\sigma}
\bmdefine{\bPsi}{\Psi}
\bmdefine{\bPi}{\Pi}
\bmdefine{\btheta}{\theta}

\newcommand{\del}{\partial}
\renewcommand{\d}{{\rm{d}}}

\newcommand{\tx}{\tilde{x}}

\newcommand{\tht}{\theta}

\newcommand{\tphi}{\tilde{\phi}}
\newcommand{\tPsi}{\tilde{\Psi}}

\newcommand{\be}{\tilde{\bm e}}

\newcommand{\bdF}{\dot{{\bm F}}}

\journal{}

\begin{document}

\begin{frontmatter}



\title{Generalized smoothed particle hydrodynamics with overset methods in total Lagrangian formulations}


\author[univ.tohoku]{Huachao Deng\corref{contrib}}
\author[ifs.tohoku]{Yoshiaki Abe$^{*,}$\corref{contrib}}
\author[univ.tohoku,univ.washington]{Tomonaga Okabe}

\cortext[corauthor]{Corresponding author}
\cortext[contrib]{Authors contributed equally}

\address[univ.tohoku]{Department of Aerospace Engineering, Tohoku University, 6-6-01, Aramaki-Aza-Aoba, Aobaku, Sendai, 980-8579, Japan}
\address[ifs.tohoku]{Institute of Fluid Science, Tohoku University, 2-1-1, Katahira, Aobaku, Sendai, Tohoku University, 980-8577, Japan}
\address[univ.washington]{Department of Materials Science and Engineering, University of Washington, BOX 352120, Seattle, WA 98195, U.S.A.}
\setstretch{1.2}
\begin{abstract}
This study proposes a generalized coordinates based smoothed particle hydrodynamics (GSPH) method with overset methods using a Total Lagrangian (TL) formulation for large deformation and crack propagation problems.
In the proposed GSPH, the physical space is decomposed into multiple domains, each of which is mapped to a local coordinate space (generalized space) to avoid coordinate singularities as well as to flexibly change the spatial resolution.
The smoothed particle hydrodynamics (SPH) particles are then non-uniformly, e.g., typically in the boundary-conforming way, distributed in the physical space while they are defined uniformly in each generalized space similarly to the normal SPH method, which are numerically related by a coordinate transformation matrix.
By solving a governing equation in each generalized space, the shape and size of the SPH kernel can be spatially changed in the physical space so that a spatial resolution is adaptively varied a priori depending on the deformation characteristics, and thus, a low-cost calculation with the less number of particles is achieved in complex shape structures.
\end{abstract}



\end{frontmatter}

\setstretch{1.3}

\section{Introduction}
Crack propagation and impact problems are often encountered in engineering applications such as for predictions of fatigue crack growth on aircraft structures and impact damage between bird and aircraft known as a bird-strike event. 
Although predicting these phenomena is significant from a fail-safe design aspect, experimental evaluations of structure damage are generally costly in time and resource, which greatly motivates the use of numerical simulations in the present field.
Traditional mesh-based numerical methods, such as a finite-element method (FEM) including the extended FEM (XFEM), effectively perform and have been widely adopted in these problems;
meanwhile, the mesh-based methods often suffer from significant drawbacks in mesh generation for complex and largely deformed boundaries due to mesh distortion and entanglement.
Smoothed particle hydrodynamics (SPH) is one of the well-established meshfree methods, which is able to overcome those drawbacks in the mesh-based methods.

The SPH has been originally developed by Gingold and Monagan~\cite{Gingold1977} and Lucy~\cite{Lucy1977} for astrophysics, which has been later extended to free-surface-flow problems by Monaghan~\cite{monaghan1995sph} and large strain solid mechanics including a crack propagation and impact problems~\cite{Libersky1993,benz1995simulations,liu2003computer}.
The other meshfree methods such as element-free Galerkin methods~\cite{Nayroles1992,Belytschko1994}, meshless local Petrov-Galerkin method~\cite{Atluri1998}, moving least squares methods, and material point method~\cite{Sulsky1994} are also seen in recent studies, wherein the reproducing kernel particle method can be particularly mentioned here, which adds a correction function in the kernel representation and improves an accuracy and efficiency especially for impact and large deformation problems~\cite{Liu1995}.
Among them, the SPH is the basis of many other meshless methods as above and has been continuously developed in many practical engineering problems.
The original SPH left a few shortcomings in its formulation, e.g., inconsistency~\cite{Vignjevic2000,Bonet2002}, tensile instability~\cite{Belytschko2000}, and rank deficiency~\cite{Swegle1995}.
The inconsistency stands for the issue in consistency even in zero order approximation with arbitrary distribution of particles. 
This issue has been resolved by so-called the corrected SPH (CSPH)~\cite{Vignjevic2000,Bonet2002}, which achieves a first-order consistency in the particle approximation and ensures the conservation of linear and angular momentum in the governing equations.
Another critical shortcoming was the tensile instability often encountered in solid mechanics problems, of which the root has been identified as the SPH formulation in the deformed current configuration, i.e., an updated-Lagrangian formulation~\cite{Belytschko2000}. 
After a number of solutions to the tensile instability~\cite{Dyka1995,Monaghan2000,Vignjevic2000,Gray2001}, a total-Lagrangian SPH (TLSPH)~\cite{Bonet2002,Vignjevic2006} has been proposed so that the governing equation including the kernel function are formulated in the total-Lagrangian framework.
Based on those backgrounds, this study will focus on the TLSPH to solve solid-mechanics problems including large deformation and crack propagation with a view to application of our techniques described later to the other meshless methods. 

In standard applications of SPH to solid mechanics, particles are initially aligned uniformly so that the interaction between any two particles can be well controlled by the kernel function that has an isotropic shape in the physical space.
Meanwhile, it is often beneficial to locally control the resolution for reducing the computational cost particularly when considering a complex geometry and crack propagation problems. 
Such non-uniform resolution methods have been extensively developed in fluid problems, wherein an adaptive particle refinement technique, i.e., particle splitting and merging based on refinement criteria, has been a main stream.
For instance, Feldman and Bonet~\cite{Feldman2007} presented a particle splitting method for adaptive refinement in dynamic fluid problems and investigated numerical errors;
Vacondio {\it et al.}~\cite{Vacondio2013} introduced a variable-resolution technique by splitting a parent particle into child particles and vise versa for refinement and coarsening, respectively.
Subsequently, a number of adaptive particle methods have been proposed on controlling the resolution in fluid problems~\cite{Barcarolo2014,Khorasanizade2015,Chiron2018,Sun2018,Yang2019}.
However, those techniques often encounter a problem in conservation of mass~\cite{Barcarolo2014} and generally require a cumbersome implementation with respect to a parallel computation.
In solid mechanics, the particle distribution is not drastically changing in time unlike fluid problems, and thus, the static refinement~\cite{Lopes2013} can be primarily considered, wherein the refinement zone with particle splitting is static and defined beforehand depending on the physical properties of computational targets~\cite{Feldman2007,Vacondio2013,Barcarolo2014,Chiron2018}. 
Such a static refinement with particle splitting is effective if the physical properties requiring a high resolution are preinformed; nevertheless, the particle splitting technique still requires some empirical criteria for changing the resolution and often leads to a complicated implementation.

Yashiro and Okabe~\cite{Yashiro2015} proposed SPH in generalized coordinate systems, i.e., generalized SPH (GSPH).
In the GSPH, particles are nonuniformly, typically in the boundary-conforming way, distributed in the physical space, wherein the governing equation is formulated in the generalized space so that the particles are uniformly distributed in the generalized space and solvable using a standard SPH with coordinate transformation matrix. 
It is noteworthy that in the GSPH, the governing equations in the generalized space are derived using a tensor analysis, which makes the coordinate transformation applicable to many terms with high-order derivatives frequently required in the solid mechanics.
They demonstrated quasi-static three-point bending of a thin plate and a high-velocity impact problem, where the intervals between particles are varied in the thickness and in-plane directions, thereby leading to a more efficient simulation compared to the standard SPH. 

Although the governing equations are formulated in the generalized space, their demonstration was limited to non-curved geometries,
and thus, the application of the GSPH to curved geometries has not yet been attempted in solid problems.
Note that in fluid problems, the similar method has been proposed by Tavakkol {\it et al.}~\cite{Tavakkol2017} as a curvilinear SPH with the coordinate transformation using the Chain-rule relation between the Cartesian and curvilinear coordinates.
Meanwhile, the use of such coordinate transformations often encounters a coordinate singularity issue:
for instance, the center axis of the cylindrical coordinate cannot be uniquely defined in the generalized space, which is often useful in solid problems such as the cylinder bending and Taylor impact problems.

This study aims at extending the original GSPH to more complex geometries including curved boundaries, wherein the tensor-analysis-based coordinate transformation~\cite{Yashiro2015} are reformulated based on the TLSPH in the solid mechanics. 
Furthermore, to overcome a drawback in the coordinate singularity as above and provide a more flexible refinement configuration, we propose the GSPH augmented by an overset method.
The GSPH with such coordinate transformations is inspired from the isoparametric analysis in the FEM in solid mechanics community, and thus, it is possible to utilize well-established knowledge on meshing strategies in the FEM community, which can be more advantageous than other refinement methods in SPH.

\section{Methodologies}\label{sec:method}
First, the coordinate transformation between the physical and generalized spaces is introduced based on the tensor analysis~\cite{MarsdenText,MalvernText}, with which the governing equations in the physical space are transformed to those in the generalized space. 
Second, the SPH approximation is described with its kernel representation and discretization using a particle approximation.
Then, an overset method is introduced to augment the present GSPH.
Finally, the governing equations in the TL formulation are described with the algorithm of the present methodology.
\subsection{Coordinate transformation}
The physical coordinate system is defined as $(\tx_1,\tx_2,\tx_3)$ $(=(\tx^1,\tx^2,\tx^3))$ with the basis of $\{\be_1,\be_2,\be_3\}$ (or $\{\be^1,\be^2,\be^3\}$).
The generalized coordinate system is defined as $(\tht^1,\tht^2,\tht^3)$ (or $(\tht_1,\tht_2,\tht_3)$) with the basis of $\{\bg_1,\bg_2,\bg_3\}$ (or $\{\bg^1,\bg^2,\bg^3\}$). 
The definition of covariant and contravariant basis $\bg_i$ and $\bg^i$ are 
\begin{align}
\bg_i = \frac{\del \tx_\alpha}{\del \tht^i}\be_\alpha,\quad
\bg^i = \frac{\del \tht^i}{\del \tx_\alpha}\be_\alpha,
\end{align}
where $\{\alpha,\beta,\gamma,\ldots\}$ and $\{i,j,k,\ldots\}$ are varied as $\{1,2,3\}$ and will be used for the physical and generalized coordinate system, respectively.
Hereinafter, the Einstein summation convention is used for those indices.
Note that the physical space is represented by the Cartesian coordinate system, and thus, the subindices $\{\alpha,\beta,\gamma,\ldots\}$ represent either covariant or contravariant component, which are the same.
The Cartesian basis can be written as
\begin{align}
\be_\alpha = \frac{\del \tht^i}{\del \tx_\alpha}\bg_i=
\be^\alpha = \frac{\del \tx_\alpha}{\del \tht^i}\bg^i.
\end{align}
We are interested in the coordinate transformation for derivatives of arbitrary physical quantities (scalar, vector, and tensor) between the Cartesian and generalized coordinate systems while keeping the Cartesian component to be differentiated.
According to the notations above, an arbitrary vector $\bphi$ is written in each coordinate system as follows:
\begin{align}
 \bphi&=\tphi_\alpha\be_\alpha\\
    &=\phi^i\bg_i=\phi_i\bg^i,
\end{align}
Let us consider the coordinate transformation for covariant derivative of an arbitrary vector $\bphi\otimes{\nabla}$ as follows:
\begin{align}
\bphi\otimes{\nabla}
&:=\left(\tphi_\alpha\be_\alpha\right) \otimes \left(\be_\beta\frac{\del }{\del \tx_\beta}\right)
=\underbrace{\left(\tphi_\alpha\be_\alpha\right)}_{\text{Cartesian basis}} \otimes \underbrace{\left(\frac{\del \tx_\beta}{\del \tht^i}\bg^i \frac{\del }{\del \tx_\beta}\right)}_{\text{Generalized basis}}\\
&=\frac{\del \tx_\beta}{\del \tht^i}\frac{\del\tphi_\alpha }{\del \tx_\beta} \left(\be_\alpha \otimes \bg^i \right)\notag
=\frac{\del\tphi_\alpha }{\del \tht^i} \left(\be_\alpha \otimes \frac{\del \tht^i}{\del \tx_\gamma}\be_\gamma \right)\\
&=\frac{\del \tht^i}{\del \tx_\gamma}\frac{\del\tphi_\alpha }{\del \tht^i} \left(\be_\alpha \otimes \be_\gamma \right),\label{eq:VectorGrad}
\end{align}
where the differential operator is expressed with the generalized basis while keeping the Cartesian component for the differentiated vector.
The coordinate transformation of ${\nabla}\cdot\bphi$ follows
\begin{align}
%
{\nabla}\cdot\bphi
&:=
\left(\be_\alpha\frac{\del }{\del \tx_\alpha} \right)\cdot (\tphi_\beta\be_\beta)
=\underbrace{\left(\bg^i\frac{\del \tx_\alpha}{\del \tht^i}\frac{\del }{\del \tx_\alpha} \right)}_{\text{Generalized basis}}\cdot \underbrace{(\tphi_\beta\be_\beta)}_{\text{Cartesian basis}}\notag\\
&=\frac{\del \tx_\alpha}{\del \tht^i}\frac{\del \tphi_\beta}{\del \tx_\alpha} \left(\bg^i\cdot\be_\beta\right)
=\frac{\del \tx_\alpha}{\del \tht^i}\frac{\del \tphi_\beta}{\del \tx_\alpha} \left(\frac{\del \tht^i}{\del \tx_\gamma}\be_\gamma\cdot\be_\beta\right)
=\frac{\del \tx_\alpha}{\del \tht^i}\frac{\del \tphi_\beta}{\del \tx_\alpha} \left(\frac{\del \tht^i}{\del \tx_\gamma}\delta_{\gamma\beta}\right)\notag\\
&=\frac{\del \tx_\alpha}{\del \tht^i}\frac{\del \tphi_\beta}{\del \tx_\alpha} \frac{\del \tht^i}{\del \tx_\beta}
=\frac{\del \tht^i}{\del \tx_\beta}\frac{\del \tphi_\beta}{\del \tht^i},\label{DivCartGen}
\end{align}
The gradient vector of a scalar $\nabla\phi$ is transformed as follows:
\begin{align}
%
\nabla\phi
&:=\left(\be_\alpha\frac{\del }{\del \tx_\alpha}\right)\phi
=\left(\frac{\del \tx_\alpha}{\del \tht^i}\bg^i\frac{\del }{\del \tx_\alpha}\right)\phi
=\underbrace{\left(\bg^i\frac{\del }{\del \tht^i}\right)}_{\text{Generalized basis}}\phi
=\frac{\del \phi}{\del \tht^i}\bg^i
=\frac{\del \phi}{\del \tht^i}\frac{\del \tht^i}{\del \tx_\alpha}\be_\alpha.\label{eq:GradScalar}
\end{align}
Similarly, an arbitrary tensor $\bPsi$ is written in each coordinate system as follows:
\begin{align}
\bPsi&=\tPsi_{\alpha\beta}\be_\alpha\otimes\be_\beta\\
&=\Psi_{ij}\bg^i\otimes\bg^j=\Psi^{ij}\bg_i\otimes\bg_j.
\end{align}
Therefore, the coordinate transformation of ${\nabla}\cdot\bPsi$ and $\bPsi\cdot{\nabla}$ follows:
\begin{align}
{\nabla}\cdot\bPsi
&:=
\left(\tPsi_{\beta\gamma}\be_\beta\otimes\be_\gamma\right)\cdot \left(\be_\alpha\frac{\del }{\del \tx_\alpha} \right)
=
\underbrace{\left(\tPsi_{\beta\gamma}\be_\beta\otimes\be_\gamma\right)}_{\text{Cartesian basis}}\cdot\underbrace{\left(\bg^i\frac{\del }{\del \tht^i} \right)}_{\text{Generalized basis}}\notag\\
&=
\frac{\del \tPsi_{\beta\gamma}}{\del \tht^i}\left((\bg^i\cdot\be_\beta)\be_\gamma\right)
=
\frac{\del \tPsi_{\beta\gamma}}{\del \tht^i}\left(\left(\frac{\del \tht^i}{\del \tx_\alpha}\be_\alpha\cdot\be_\beta\right)\be_\gamma\right)\notag\\
&=
\frac{\del \tPsi_{\beta\gamma}}{\del \tht^i}\left(\left(\frac{\del \tht^i}{\del \tx_\alpha}\delta_{\alpha\beta}\right)\be_\gamma\right)
=
\frac{\del \tPsi_{\beta\gamma}}{\del \tht^i}\frac{\del \tht^i}{\del \tx_\beta}\be_\gamma,\\
\bPsi\cdot{\nabla}
&:=
\left(\tPsi_{\beta\gamma}\be_\beta\otimes\be_\gamma\right)\cdot \left(\be_\alpha\frac{\del }{\del \tx_\alpha} \right)
=
\underbrace{\left(\tPsi_{\beta\gamma}\be_\beta\otimes\be_\gamma\right)}_{\text{Cartesian basis}}\cdot \underbrace{\left(\bg^i\frac{\del }{\del \tht^i} \right)}_{\text{Generalized basis}}\notag\\
&=
\frac{\del \tPsi_{\beta\gamma}}{\del \tht^i}\left((\bg^i\cdot\be_\gamma)\be_\beta\right)
=
\frac{\del \tPsi_{\beta\gamma}}{\del \tht^i}\left(\left(\frac{\del \tht^i}{\del \tx_\alpha}\be_\alpha\cdot\be_\gamma\right)\be_\beta\right)\notag\\
&=
\frac{\del \tPsi_{\beta\gamma}}{\del \tht^i}\left(\left(\frac{\del \tht^i}{\del \tx_\alpha}\delta_{\gamma\alpha}\right)\be_\beta\right)
=
\frac{\del \tPsi_{\beta\gamma}}{\del \tht^i}\frac{\del \tht^i}{\del \tx_\gamma}\be_\beta.
\end{align}
Finally, an inner product of two tensor products given by arbitrary vectors $\ba$, $\bb$, $\bc$, and $\bd$ is defined as follows:
\begin{align}
(\ba\otimes\bb)\colon(\bc\otimes\bd):=(\ba\cdot\bc)(\bb\cdot\bd).
\end{align}
Therefore, an inner product of arbitrary tensor products, $\bA\colon\bB$, is calculated as:
\begin{align}
\bA\colon\bB=
\left(\tilde{A}_{\alpha\beta}(\be_\alpha\otimes\be_\beta)\right)\colon
\left(\tilde{B}_{\gamma\delta}(\be_\gamma\otimes\be_\delta)\right)
=
\tilde{A}_{\alpha\beta}
\tilde{B}_{\gamma\delta}\delta_{\alpha\gamma}\delta_{\beta\delta}
=\tilde{A}_{\alpha\beta}\tilde{B}_{\alpha\beta}.\label{eq:TensorInner}
\end{align}
As such, all the derivatives of vectors and tensors in Eqs.~\eqref{eq:VectorGrad}-\eqref{eq:TensorInner} are represented by derivatives of the Cartesian component with respect to the generalized coordinate system. This enables us to avoid the Christoffel symbol which often becomes a burden on formulation and computation due to its complexity and high computational cost.
 For example, the gradient of a scalar given in Eq.~\eqref{eq:GradScalar} performs as follows:
first, the coordinate transformation matrix, ${\del \tht^i}/{\del \tx_\alpha}$, are computed;
second, the derivative of the scalar in generalized coordinate system, ${\del \phi}/{\del \tht^i}$, is computed by the SPH approximation in the next section;
finally, both of the terms above are multiplied, which gives the derivative of a scalar in the physical space.
The present formulation ensures that the number of particles inside each support is sufficient as long as the generalized coordinate system is defined so that the shape of each support region is appropriately deformed in the physical space. We will describe procedures and guidelines to define the generalized coordinate system later on. It should be noted that the present transformation eventually corresponds to a Chain-rule transformation of the Cartesian component, which has been generally adopted as the body-fitted coordinate system in the mesh-based schemes for computational fluid dynamics~\cite{Abe2013b,Abe2015} and also as the isoparametric formulation in the finite element analysis. 
Nevertheless, the transformation above are purely derived from the definitions of covariant and contravariant basis instead of relying on only the Chain-rule transformation, which focuses merely on components and ignores the existence of basis, and thus, the present formulation revisits and provides a rigorous derivation in terms of vector and tensor analyses. Such a tensor-analysis-based coordinate transformation has been adopted as the generalized coordinate SPH for solid dynamics in Yashiro and Okabe~\cite{Yashiro2015} without curved coordinates. 
This study provides a straightforward extension of the tensor-analysis-based generalized SPH by Yashiro and Okabe~\cite{Yashiro2015} to the curved coordinates in solid dynamics. Furthermore, the proposed tensor-analysis-based generalized SPH is augmented by an overset methodology to introduce a local coordinate and deal with singularities.

\subsection{SPH approximation}
Based on a set of the physical and generalized coordinates defined in the previous subsection, an arbitrary location $\bx=(\tx_1,\tx_2,\tx_3)$ in the physical space can be mapped to the generalized space as $\btheta=(\tht^1(\bx),\tht^2(\bx),\tht^3(\bx))$.
The GSPH performs a particle approximation in the generalized space, and thus, a physical quantity $f$ at an arbitrary point $\btheta$ in the computational domain $\Omega$ can be represented as:
\begin{align}
f(\btheta) =\int_{\Omega}f(\btheta')W(\btheta-\btheta',h)\d\btheta',\label{eq:0}
\end{align}
where a smoothing parameter $h$ represents a radius of a radially symmetric compact support.
$W$ is so-called the kernel function, and a cubic B spline function is adopted in this study as follows:
\begin{align}
W(\btheta-\btheta',h)=\alpha_D
\begin{cases}
&\displaystyle 1-\frac{3}{2}q^2+\frac{3}{4}q^3 \quad \text{if}\quad 0\leq q< 1\\
&\displaystyle \frac{1}{4}(2-q)^3 \quad \text{if}\quad 1\leq q< 2\\
&0 \quad \text{otherwise}
\end{cases}
\end{align}
where $\alpha_D=1/(\pi h^3)$ and $q=|\btheta-\btheta'|/h$.

Suppose that $N$ particles exist in the compact support of the kernel function around the particle $a$ at $\btheta=\btheta_a$, Eq.~\eqref{eq:0} for the particle $a$ is approximated as follows:
\begin{align}
f(\btheta_a)\approx \sum_{b=1}^N\frac{m_b}{\rho_b}f(\btheta_b)W_{ab},
\end{align}
where $W_{ab}=W(\btheta_a-\btheta_b,h)$, and $m_b$ and $\rho_b$ represent the mass and density of the particle $b$, respectively. 
In the rest of this paper, the subindices $a$ and $b$ represent values of the particle $a$ and $b$, with which the Einstein summation convention is not taken.
The derivative of $f$ in the $\theta^i$ direction at the particle $a$ can be approximated in two forms as follows:
\begin{align}
\left(\frac{\del f(\btheta)}{\del \theta^i}\right)_{a}
&\approx \rho_a\sum_{b=1}^Nm_b\left\{\frac{f(\btheta_a)}{\rho_a^2}+\frac{f(\btheta_b)}{\rho_b^2}\right\}\frac{\del W_{ab}}{\del \theta^i_{a}},\label{eq:SPHderiv1}\\
\left(\frac{\del f(\btheta)}{\del \theta^i}\right)_{a}
&\approx \frac{1}{\rho_a}\sum_{b=1}^Nm_b\left\{f(\btheta_a)-f(\btheta_b)\right\}\frac{\del W_{ab}}{\del \theta^i_{a}}.\label{eq:SPHderiv2}
\end{align}
%
Furthermore, the present study adopts the CSPM~\cite{chen1999corrective} to improve the consistency of particle approximation near the boundary to avoid inconsistent representation of the spatial derivatives due to a truncation of the kernel function at the boundary.

\subsection{Total Lagrangian GSPH}
\subsubsection{Discretization of governing equations}
In the total Lagrangian framework, the current position of a material point is written as $\bx=x_\alpha\be_\alpha$, and the reference position of the same point is expressed as $\bX=X_\alpha\be_\alpha$, where $\alpha=\{1,2,3\}$ stands for the index of the physical coordinate. $(x_1,x_2,x_3)$ and $(X_1,X_2,X_3)$ represent the physical coordinate of the material point in the current and reference position, respectively.
The deformation gradient $\bF$ and its rate $\dot{{\bm F}}$ are defined as follows:
\begin{align}
\bF&=\frac{\del \bx}{\del \bX}\\
&=\frac{\del x_\alpha}{\del \theta^i}\frac{\del \theta^i}{\del X_\beta}\be_\alpha\otimes\be_\beta,\\
\dot{{\bm F}}&=\frac{\del \bv}{\del \bX}\\
&=\frac{\del v_\alpha}{\del \theta^i}\frac{\del \theta^i}{\del X_\beta}\be_\alpha\otimes\be_\beta.
\end{align}
The governing equations are established on the reference configuration, which are written as:
\begin{align}
\rho&=J^{-1}\rho_0,\label{eq:gov1org}\\
\frac{\d \bv}{\d t}&=\frac{1}{\rho_0}\nabla_0\cdot\bP\\
&=\frac{1}{\rho_0}\frac{\del P_{\alpha\beta}}{\del \theta^i}\frac{\del \theta^i}{\del X_\alpha}\be_\beta,\label{eq:gov2org}
\end{align}
where $\rho$ and $\bv$ are the density and velocity, respectively, and the subscript $0$ denotes the values on the reference configuration. $J=\text{det}|\bF|$ represents the determinant of the deformation gradient $\bF$.
$\bP$ is the first Piola Kirchhoff stress tensor defined as:
\begin{align}
\bP=P_{\alpha\beta}\be_\alpha\otimes\be_\beta=J\bF^{-1}\cdot\bsigma,
\end{align}
where $\bsigma$ is the Cauchy tensor, which will be represented later in Eq.~\eqref{eq:CauchyTensor}.
In the GSPH, the governing equations are Eqs.~\eqref{eq:gov1org} and \eqref{eq:gov2org}, where the stress tensor is expressed in the physical coordinate of the reference configuration $X_\alpha$ while the spatial derivatives are taken with respect to the generalized coordinate $\theta^i$.

Based on the SPH approximation, a semi-discrete form of the governing equations Eqs.~\eqref{eq:gov1org} and \eqref{eq:gov2org} in the GSPH at the particle $a$ are as follows:
\begin{align}
\rho_a&=J^{-1}_a\rho_{0;a},\\
\frac{\d\bv_a}{\d t}
&=\rho_{0;a}\left\{\sum_{b=1}^{N}m_{0;b}\left(
\frac{P_{\alpha\beta;a}}{\rho_{0;a}^2}+\frac{P_{\alpha\beta;b}}{\rho_{0;b}^2}-\Pi_{ab;\alpha\beta}\right)
\frac{\del W_{ab}}{\del \theta_a^i}\right\}
\left(\frac{\del \theta^i}{\del X_\alpha}\right)_{a}\be_\beta.\label{eq:dvdt}
\end{align}
$\Pi_{ab;\alpha\beta}$ represents the $\alpha\beta$ components of the artificial viscosity in the physical space as follows
\begin{align}
\Pi_{ab;\alpha\beta}&=J(\bF^{-1})_{\alpha\beta}\pi_{ab},\\
\pi_{ab}=&
\begin{cases}
  \displaystyle\frac{-\beta_1\overline{c}_{ab}\phi_{ab}+\beta_2 \phi_{ab}^2}{\overline{\rho}_{ab}} \quad&\text{if}\quad \bv_{ab}\cdot\bx_{ab}<0\\
0\quad&\text{if}\quad \bv_{ab}\cdot\bx_{ab}\geq 0,
\end{cases}
\end{align}
with the definitions of
\begin{align}
\phi_{ab}&=\frac{h_{ab}(\bv\cdot\bx_{ab})}{(|\bx_{ab}|^2+\varphi^2)},\quad
\varphi=0.01h^2,\quad
\overline{c}_{ab}=0.5(c_a+c_b),\\
\overline{\rho}_{ab}&=0.5(\rho_a+\rho_b),\quad
h_{ab}=0.5(h_a+h_b),\quad
\bv_{ab}=\bv_a-\bv_b,\\
\bx_{ab}&=\bx_a-\bx_b,
\end{align}
where $c_{a}$ is the speed of sound at the particle $a$.
The SPH approximations of the deformation gradient $\bF$ and its rate $\bdF$ at the particle $a$ are
\begin{align}
\bF
&=\left\{ \frac{1}{\rho_{0;a}}\sum_{b=1}^{N}m_{0;b}
\left(x_{\alpha;a}-x_{\alpha;b}\right)
\frac{\del W_{ab}}{\del \theta^i_a}\right\}
\left(\frac{\del \theta^i}{\del X_\beta}\right)_{a}\be_\alpha\otimes\be_\beta,\label{eq:F}\\
\bdF
&=\left\{ \frac{1}{\rho_{0;a}}\sum_{b=1}^{N}m_{0;b}
\left(v_{\alpha;a}-v_{\alpha;b}\right)
\frac{\del W_{ab}}{\del \theta^i_a}\right\}
\left(\frac{\del \theta^i}{\del X_\beta}\right)_{a}\be_\alpha\otimes\be_\beta.\label{eq:dF}
\end{align}
Equations~\eqref{eq:dvdt}, \eqref{eq:F}, and \eqref{eq:dF} contain $\del\theta^i/\del X_\alpha$ for the coordinate transformation between the physical and the generalized coordinates of the reference frame.
The coordinate transformation matrix at the particle $a$ is computed as
\begin{align}
\left(\frac{\del X_\alpha}{\del \theta^i}\right)_a 
=\left\{\sum_{b=1}^N m_{0;b}\left(X_{\alpha;b}-X_{\alpha;a}\right)\frac{\del W_{ab}}{\del \theta^i_a}\right\} 
.\label{eq:met}
\end{align}
$\del\theta^i/\del X_\alpha$ at the particle $a$ can be computed as the inverse of the coordinate transformation matrix given by Eq.~\eqref{eq:met}.

\subsubsection{Constitutive equations}
The Jaumann stress rate is used in the present study as follows:
\begin{align}
\dot{\bsigma}=\bsigma^{\nabla}+\bsigma\cdot\bW^T+\bW\cdot\bsigma,\label{eq:CauchyTensor}
\end{align}
where $\bW$ is the spin tensor defined as:
\begin{align}
\bW&=\frac{1}{2}\left(\bL-\bL^T\right),\\
\bL&=\frac{\del\bv}{\del\bx}=\frac{\del\bv}{\del\bX}\frac{\del\bX}{\del\bx}=\dot{\bF}\bF^{-1}.
\end{align}

\subsubsection{Rankine criterion}
This study demonstrates three-dimensional crack propagation problems, where the Rankine criterion is used to deal with the brittle crack propagation~\cite{islam2019total}.
The interaction factor $f_{ab}$ based on the damage index $D_{ab}$ is defined to characterize the interaction state of particle $a$ and $b$ as
\begin{align}
f_{ab}=1-D_{ab}.
\end{align}
The damage evolution~\cite{islam2019total} is based on the following criterion:
\begin{align}
D_{ab}=\begin{cases}
1 & \text{if}\quad \displaystyle{\frac{(r_{ab})_t-(r_{ab})_0}{(r_{ab})_0}}\geq \varepsilon_\text{max}\\
0 & \text{otheriwse},
\end{cases}
\end{align}
where $\varepsilon$ is chosen to be 0.03;
$(r_{ab})_t$ and $(r_{ab})_0$ are the distances between particles $a$ and $b$ at the current and reference configurations, respectively.
In the beginning of the simulation, there is no damage in the model, i.e., $D_{ab}=0$ and $f_{ab}=1$, for all of the interaction pairs.
When damage initiates, i.e., $D_{ab}=1$ and $f_{ab}=0$, the interaction between particles $a$ and $b$ is removed, and a crack surface is generated implicitly.
This criterion models a brittle failure if the distance between two particles is greater than a threshold.

\subsubsection{Thermo-visco-plastic behavior}
In this study, Johnson-Cook model is utilized to consider plastic hardening, rate dependency, and thermal softening~\cite{Islam2020}.
The yield stress $\sigma_y$ is expressed as
\begin{align}
\sigma_y=(A+B\overline{\varepsilon}_{pl}^n)(1+C\log \dot{\overline{\varepsilon}}_{pl}^*)(1-T^{*m}),
\end{align}
where $A$ is an initial yield stress of the material;
$B$, $C$, $m$, and $n$ are the hardening parameters.
$\overline{\varepsilon}_{pl}^n$ is a dimensionless effective plastic strain, and $\dot{\overline{\varepsilon}}_{pl}^*$ is defined as
\begin{align}
\dot{\overline{\varepsilon}}_{pl}^*=\frac{\overline{\varepsilon}_{pl}}{\overline{\varepsilon}_{0}},
\end{align}
where $\overline{\varepsilon}_{pl}$ and $\overline{\varepsilon}_{0}$ are the effective plastic strain rate and reference strain rate, respectively.
Nondimensional temperature is defined as 
\begin{align}
T^*=\frac{T-T_r}{T_m-T_r},
\end{align}
where $T$, $T_r$, and $T_m$ are the current, room, and melting temperature of the material, respectively.
The increase of temperature is caused by a plastic work as follows:
\begin{align}
\Delta T=\chi\frac{\Delta w_p}{\rho C_p},
\end{align}
where $\Delta w_p$, $\rho$, $C_p$, and $\chi$ are increment of a plastic work, density, specific heat capacity, and empirical constant $\chi=0.9$.

The Von Mises yield criterion $y_f=\sqrt{3J_2}-\sigma_y$ is adopted to determine if the stress state beyond the yield surface, where $J_2=\bS:\bS/2$ is a second invariant of deviatoric stress tensor $\bS$.
The Wilkins criterion $\bS_n=c_f\bS$ is used for a return mapping when the trial elastic stress state exceeds the yield surface, where $c_f=\min(\sigma_y/\sqrt{3J_2},1)$, and $\bS$ is the corrected deviatoric stress tensor.
Finally, the following equations are used to compute the increment of plastic strain, the increment of effective plastic strain, and the accumulated plastic work density~\cite{Islam2020} as follows:
\begin{align}
\Delta {\bm \varepsilon_{pl}}&=\frac{1-c_f}{2G}\bS,\\
\Delta \overline{\varepsilon}_{pl}&=\sqrt{\Delta {\bm \varepsilon_{pl}}:{\bm \varepsilon_{pl}}}=\frac{1-c_f}{3G}\sqrt{\frac{3}{2}\bS:\bS},\\
\Delta w_p&={\bm \varepsilon_{pl}}:\bS_n.
\end{align}

\subsubsection{Damage model}\label{sec:damagemodel}
A correct damage model will be employed to deal with a fracture in steel-plate penetration problems in this study.
The Johnson-Cook model coupled with a damage model is adopted with the modified yield stress~\cite{Borvik1999} as follows:
\begin{align}
\sigma_y=(1-D)(A+Br^n)(1+C\log{\dot{r}^*})(1-T^{*m}),
\end{align}
in which $r$ is the damage accumulated plastic strain given as $\dot{r}=(1-D)\dot{\overline{\varepsilon_{pl}}}$.
$D$ is determined by the Johnson-Cook criterion as
\begin{align}
D=\sum\frac{\Delta\overline{\varepsilon_{pl}}}{\varepsilon_f},
\end{align}
where $\Delta\overline{\varepsilon_{pl}}$ and $\varepsilon_{f}$ are the incremental effective plastic and fracture strain, respectively.
$\varepsilon_f$ is calculated as follows:
\begin{align}
\varepsilon_f=[D_1+D_2\exp(D_3\sigma^*)][1+\dot{\overline{\varepsilon_{pl}}}^*]^{D_4}[1+D_5T^*],
\end{align}
where $\sigma^*=\sigma_m/\sigma_{eq}$ is a stress tri-axiality ratio, and $\sigma_m$ is a mean stress.
$D_i$ ($i=1,\ldots,5$) are material constants as $D_1=0.0705$, $D_2=1.732$, $D_3=-0.54$, $D_4=-0.015$, and $D_5=0.0$.


\subsection{Overset method}
\begin{figure}
\centering
\includegraphics[width=1\textwidth]{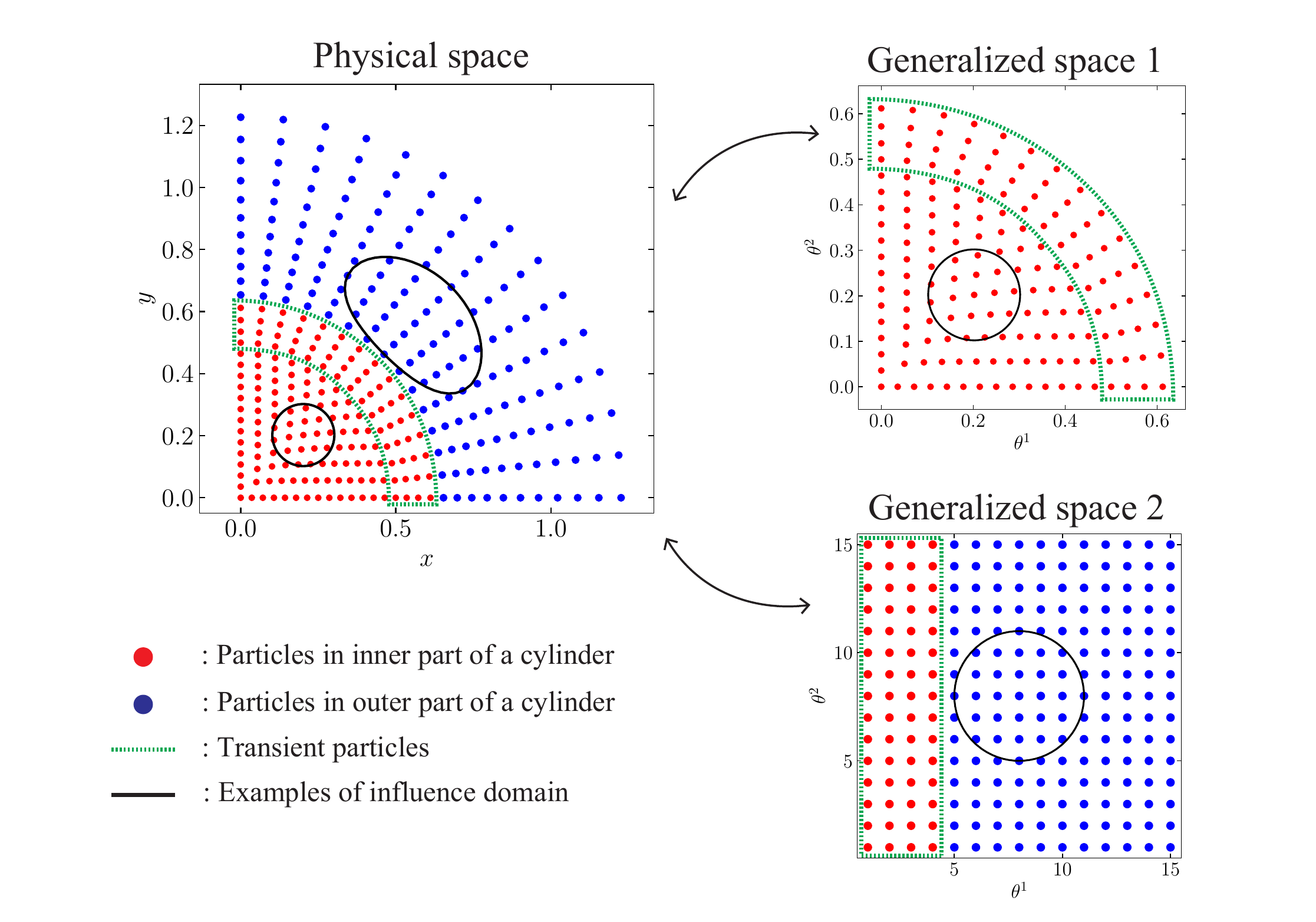}
\caption{Schematic of particle distributions in the GSPH with a overset method}
\label{fig:overset}
\end{figure}

Based on the coordinate transformation between the physical and generalized spaces, the GSPH is able to deal with nonuniform particle distributions while keeping the standard SPH discretization in the generalized space for uniform particle distributions.
However, such a coordinate transformation often suffers from coordinate singularities, e.g., singularity on an axis of the cylindrical coordinate, which limits the applicability of the GSPH to further complex geometries.
We propose the GSPH with an overset method to overcome this issue. The schematic of the overset method is illustrated in Fig.~\ref{fig:overset}.
Let us consider a two-dimensional quarter cylinder as an example.
If the cylindrical coordinate is adopted as an generalized coordinate in the entire domain, the particles in the vicinity of the origin $(x,y)=(0.0,0.0)$ shows a highly-dense distribution, and the particle at the origin of the physical space cannot be uniquely mapped to the generalized space.
Therefore, in Fig.~\ref{fig:overset}, the computational domain is decomposed into two subdomains of inner and outer parts of the cylinder, where the red and blue particles are mapped to the generalized space 1 and 2, respectively~\footnote{Some of the red particles are defined as transient particles and are mapped to the generalized space 2 as well, which will be explained shortly.}.
The cylindrical coordinate is adopted to the blue particles, which are defined as the rectangular domain in the generalized space 2. 
Meanwhile, the distribution of the red particles are the same between the physical and generalized space 1, and thus, the coordinate singularity is removed at the origin $(x,y)=(0.0,0.0)$.
The calculations of Eq.~\eqref{eq:dvdt} at red and blue particles are carried out in each generalized space with the corresponding coordinate transformation matrix. 
For the communication between two generalized spaces, transient particles are defined in the green dashed line, which are mapped to both of the generalized spaces 1 and 2.
If the interaction pair which is identified as $a$ and $b$ of Eqs.~\eqref{eq:dvdt} and \eqref{eq:met} includes red particles, the kernel is defined in the generalized space 1.
In this study, it is assumed that the generalized space 1 is more comprehensive in that the particles related to the generalized space 2 (cylindrical coordinate) can be also mapped to the generalized space 1, but not vice versa.

Finally, black solid lines in Fig.~\ref{fig:overset} indicates examples of influence domain, which is also called a kernel shape in this study, in the physical and generalized spaces. The influence domain is defined as a circle around a particle in the generalized space for this two-dimensional example. The shape of the influence domain in the generalized space 2 is deformed according to the distribution of the blue particles in the physical space. Meanwhile, the shape of the influence domain is kept as a circle in the physical space for the red particles as the generalized space 1 is the same as the physical space in this example. As such, the shape of the influence domain can be varied so that the resolution of the particles adaptively and efficiently changes depending on the shape of the geometry and associated particle distributions.

\subsection{Algorithm}
As a summary, Algorithm~\ref{alg:1} shows the computational process of the GSPH in the TL formulation with an overset method.
The velocity-verlet~\cite{Wilson1982,LSDYNA} method is adopted for the time integration in this study. In the Algorithm~\ref{alg:1}, the subscripts $t$, $t+\Delta t$, and $t+\Delta t/2$ indicate the time of each quantity.
\begin{figure}[!t]
\begin{algorithm}[H]
\setstretch{1.2}
\caption{GSPH in the TL formulation with an overset method}\label{alg:1}
\begin{algorithmic}[1]
\STATE Define initial conditions including a particle distribution $\bX$ and $\btheta$ in the physical and generalized spaces, respectively;
\IF{the number of generalized spaces $\geq 2$} 
\STATE Define transient particles with two neighbouring generalized coordinates;
\ENDIF
\STATE Find interaction particles in the influence domain and calculate the derivative of the kernel function $\del W/\del \theta^i$ in each generalized space;
\STATE Calculate the coordinate transformation matrix between the physical and generalized coordinates and store $\del \theta^i/\del X_\alpha$;
\FOR{$n=1$ to $n_{\text{end}}$} 
\STATE Define the timing to be updated as $t+\Delta t= n\Delta t$
\STATE Calculate velocity at $t+\Delta t/2$ as $\bv_{t+\Delta t/2}=\bv_t+\dot{\bv}_t\Delta t$/2;
\STATE Update the current position of particles at $t+\Delta t$ as $\bx_{t+\Delta t}=\bx_t+\bv_{t+\Delta t/2}\Delta t$;
\STATE Calculate the deformation gradient $\bF$ at $t+\Delta t$;
\STATE Update the Cauchy stress $\bsigma$ and the first Piola Kirchhoff stress $\bP=J\bF^{-1}\cdot\bsigma$ at $t+\Delta t$;
\STATE Calculate the acceleration $\ddot{\bu}_{t+\Delta t}$ of particles at $t+\Delta t$;
\STATE Calculate the velocity of particles at $t+\Delta t$ as $\bv_{t+\Delta t}=\bv_{t+\Delta t/2}+\dot{\bv}_{t+\Delta t}\Delta t$/2;
\STATE Output the field variables at $t+\Delta t$.
\ENDFOR
\end{algorithmic}
\end{algorithm}
\end{figure}
%
%


\section{Conclusions}\label{sec:conclusion}
This study has proposed the GSPH using an overset method and the TL formulation in solid mechanics, in which the coordinate transformation technique between the physical and generalized (local coordinate) spaces is utilized to control a spatial resolution and reduce the number of SPH particles.
The main conclusions are summarized as follows.
\begin{itemize}
\item[1] The proposed GSPH allows non-uniform particle distributions in the physical space to locally vary the spatial resolution, while the governing equations are solved in the generalized space with uniform particle distributions. The coordinate transformation representation has been extended from the original GSPH~\cite{Yashiro2015} to curved coordinates using tensor analysis techniques. Furthermore, this is for the first time to generally represent the vector and tensor formulations with their gradients in the SPH discretization on curved coordinate systems, which is useful to implement higher-order tensors in more general solid-mechanics problems.
\item[2] The proposed GSPH is augmented by the overset method, wherein the computational domain is decomposed into multiple subdomains with local coordinates. The overset method enables us to control the spatial resolution more flexibly: for example, a non-contact zone in ballistic penetration of steel plates problem can contain the less number of particles compared to that in the contact zone. Furthermore, coordinate singularities, which is inevitable in the cylindrical or spherical coordinate system, can be avoided by combining the singularity-free coordinate system in the overset framework.
\item[3] To alleviate the tensile instability, the TL formulations have been implemented for the finite deformation problems, which is for the first time in the SPH formulation with coordinate transformation techniques. 
\item [5] The proposed GSPH discretizes the governing equations in the generalized space, where the SPH particles are uniformly distributed as in the standard SPH. Therefore, the GSPH code can be developed from existent SPH codes without a cumbersome process by implementing the coordinate transformation matrix between the physical and generalized spaces.
\end{itemize}
This is for the first time to formulate the SPH in the generalized coordinate system with the TL framework as well as with the overset configuration in solid mechanics, which enables a more flexible static resolution control and stable simulations.
Several challenging numerical tests are now being performed and have already provide positive results, which will be updated in the manuscript later.

\section*{Acknowledgements}
    This work was supported by Council for Science, Technology and Innovation(CSTI), Cross-ministerial Strategic Innovation Promotion Program (SIP), “Materials Integration” for revolutionary design system of structural materials (Funding agency: JST).


%
 \bibliographystyle{elsarticle-num} 


\bibliography{./references/sph}



\end{document}